\documentstyle[manuscript,aps,epsf]{revtex}
\newcommand{\be}{\begin{equation}}
\newcommand{\ee}{\end{equation}}
\newcommand{\ba}{\begin{eqnarray}}
\newcommand{\ea}{\end{eqnarray}}
\newcommand{\ci}[1]{\cite{#1}}
\newcommand{\lab}[1]{\label{#1}}
\begin{document}
\draft
\title{ Additional ways to determination of structure \\
        of  high energy elastic scattering amplitude}


\author{O. \ V. \ Selyugin$^{1}$ \thanks{selugin@thsun1.jinr.ru}}

\address{\it Bogoliubov Laboratory of Theoretical Physics, \\
Joint Institute for Nuclear Research,
141980 Dubna, Moscow region, Russia}

\maketitle

\abstract
 {
   Several methods of  extracting the magnitude of the
   real parts of the elastic hadron scattering amplitude -
  from the experimental data are presented.
  These methods allows us to obtain
  the real parts at one definite point of the transfer momentum and with
  a high accuracy without  knowledge of the precise values of the
  normalization coefficient and with minimum  theoretical assumptions.
}


     The obtaining of the structure of the hadron scattering amplitude
 is an important task both for a theory and an experiment.
 QCD has a certain difficulty in  calculating
 the magnitude of the scattering amplitude in the diffraction range,
  as well as  its phase and the energy dependence
 in the diffraction range.
 For our  deeper understanding, the work of
 such a fundamental relation as dispersion and
 local dispersion relations  requires the knowledge of the structure of the
 scattering amplitude with an accuracy, we can obtain \cite{mart}.
 Also note that in \ci{kh} it was shown  that the knowledge of the behavior
 of $\rho$ - the    ratio of the real to imaginary part of
   the spin-non-flip amplitude
  can be used for checking  the local QFT  already in the LHC energy region.

      A large number of experimental and theoretical studies of the
 high energy elastic
 proton-proton  and proton-antiproton scattering at  small angles
 gives a rich information about this  process,
 which allows to narrow the circle of examined models
 and  to  put a number of the difficult  problems
 which are not  solved  entirely in  the meanwhile.
 Especially this concerns  the  energy  dependence  of  a number
 of characteristics of these reactions and the contribution as the odderon.

      A great deal of these questions are connected with the dependence
 of the spin-non-flip phase of hadron-hadron scattering with s and t.
 The most of the models define  the  real  part  of  the  scattering
 amplitude phenomenologically. Some  models  used  the  local
 dispersion relations       and   the   hypothesis   of   the   geometrical
 scaling. As is known, using some  simplifying assumption, the
 information about   the   phase   of
 the   scattering   amplitude   can   be
 obtained from the experimental data at small momentum transfers   where
 the interference of the electromagnetic and hadron amplitudes takes  place.
 On the whole, the obtained information confirms the local dispersion
 relations.
   It was shown in \ci{k-l} that
 the uncontradicted description of the experimental data in   the range
 of ISR and SPS  can be obtained in the case of a rapidly  changing
 phase when the real part of the scattering amplitude grows quickly   in
 the range of small t   and   becomes dominant.

     The standard procedure to extract the magnitude of the real part
  includes the fit of the experimental data taking the magnitude
   of total cross section, slope, $\rho$, and, sometimes the normalization
  coefficient as free parameters:
\ba
   \sum_{i}^{k} \frac{(n d\sigma_{i}/dt_{exp}-d\sigma_{i}/dt)^2}{\Delta_{exp.}^2}
\ea
where  $d\sigma_{i}/dt_{exp}$  is the differential cross sections
 at point $t_i$  with the statistical error $\Delta_{i}$
extracted from
 the measured $dN/dt$ using, for example, the magnitude of luminosity.
The procedure require a sufficiently wide interval of $t$ and large number
 of experimental points.

   The theoretical representation of the differential cross-sections is
\ba
{d\sigma\over dt}&= 2 \pi
[\vert \Phi_1\vert^2+\vert \Phi_2\vert^2+ \vert \Phi_3\vert^2+\vert \Phi_4\vert^2+4\vert
 \Phi_5\vert^2]\ ,
\ea

The total helicity amplitudes can be written as a sum of nuclear $\Phi_i^h(s,t)$
and electromagnetic  $\Phi_i^e(s,t)$ amplitudes :
\ba
\Phi_i(s,t)=\Phi_i^h(s,t)+e^{i\alpha\varphi}\Phi_i^e(s,t)\ .
\ea
where $\Phi_i^e(s,t)$ are the leading-terms at high energies of the one-photon
amplitudes as defined, for example,
 in \ci{leader} and the common phase $\varphi$,     is
\ba
\varphi=-[\gamma+\log\big(B(s,t)\vert t\vert/2\big)+\nu_1+\nu_2]
\ea
where $B(s,t)$ is the slope of the nuclear amplitude, and $\nu_1$ and $\nu_2$ are
small correcting terms  define the behavior of the Coulomb-hadron phase
at small momentum transfers (see, \ci{selprd}).
At very small t and fixed s, these electromagnetic amplitudes are such that
$
\Phi_1^e(s,t) = \Phi_3^e(s,t)\sim\alpha/t \ ,
\Phi_2^e(s,t) = -\Phi_4^e(s,t)\sim \alpha\cdot \hbox{const.}\ ,
\Phi_5^e(s,t)  \sim  -\alpha/\sqrt{\vert t\vert}\ .
$
  We assume, as usual, that at high energies and small angles
  the double-flip amplitudes are small with respect to the spin-nonflip one
  and that spin-nonflip amplitudes are approximately equal. Consequently,
  the observables  are determined by two amplitudes:
  $ F (s,t) = \Phi_{1}(s,t) + \Phi_{3}(s,t)$ and $F_{sf}(s,t) = \Phi_{5}(s,t)$.
  So,
\ba
d\sigma/dt &= \pi [ (F_{C} (t))^2
          + (Re F_{N}(s,t))^{2}+ (Im F_{N}(s,t))^{2})      \nonumber \\
 &+ 2 (\rho(s,t)+ \alpha \varphi(t)) F_{C}(t) Im F_{N}(s,t)] . \label{ds2}
\ea
 $F_{C}(t) = \mp 2 \alpha G^{2}/|t|$ is the Coulomb amplitude;
$\alpha$ is the fine-structure constant  and $G^{2}(t)$ is  the  proton
electromagnetic form factor squared;
$Re\ F_{N}(s,t)$ and $ Im\ F_{N}(s,t)$ are the real and
imaginary parts of the nuclear amplitude;
$\rho(s,t) = Re \ F_{N}(s,t) / Im \ F_{C}(s,t)$.
Just this formula is used for the fit  of  experimental  data
determining the Coulomb and hadron amplitudes and the Coulomb-hadron
phase to obtain the value of $\rho(s,t)$.

   Numerous  discussions of the value of $\rho (s,t)$
 measured by
 the UA4 \ci{ua4} and UA4/2 \ci{ua42} Collaborations at $\sqrt{s}=541$ GeV have revealed
 the ambiguity in the definition of this parameter \ci{selpl},
 and, as a result, it has
 been shown that one has some trouble in extracting, from experiment,
  the total cross sections
 and the value of the real parts of the scattering amplitudes \ci{gns}.
 In fact, the problem is that we have at
  our disposal only one observable $d\sigma/dt$
 for two unknowns  the real part and imaginary part of the hadron
  spin-non-flip amplitude.
 So, we need either some additional experimental information which
 would allow us to determine independently the real and imaginary parts of
 the nonflip hadron elastic scattering amplitude or develop some new ways
  to determine the magnitude of the phase of scattering amplitude
  with minimum theoretical assumptions.
          One of the most important points in the definition
  of  the real part of scattering amplitude
  is the knowledge of the normalization coefficient and
  the magnitude of $\sigma_{tot}(s)$.

     To obtain the magnitude of   $Re F_{N}(s,t)$,
  we fit the differential cross sections either taking into account
 the value of $\sigma_{tot}$ from another experiment, to decrease the
 errors,  as made
 by the UA4/2 Collaboration, or taking  $\sigma_{tot}$ as a free
 parameter, as made in \ci{selpl}.
 If one does not take the normalization coefficient as  a free parameter in
 the fitting procedure, its definition requires the knowledge of
 the behavior of imaginary and real parts of the scattering amplitude
 in the range of small transfer momenta and the magnitude of
 $\sigma_{tot}(s)$ and $\rho(s,t)$.

     Note three points. First, in any case, we should take into account
 the errors in $\sigma_{tot}(s)$. Second, this method means that
 the imaginary part slope of the scattering amplitude equals the
 slope of its real part in  the examined range of transfer momenta,
 and for the best fit, one should take the interval of transfer momenta
  sufficiently large.
 Third, the magnitude of $\rho(s,t)$ thus obtained  corresponds to
 the whole interval of transfer momenta.

     In this report, we briefly describe  some new procedures of simplifying
 the determination of elastic scattering amplitude parameters.

   From equation   (\ref{ds2})
    one can obtain the equation for  $Re F_{N}(s,t)$
   for every experimental point - $i$
\ba
  && Re F_{N}(s,t_i)= -Re F_{C}(s,t)   \nonumber    \\
  & & \pm [n / \pi d\sigma/ dt(t=t_i)
   - (Im F_{C}(t_i)+Im F_{N}(t_i))^2]^{1/2}.
                                               \label{rsq}
\ea

 As the imaginary part of scattering amplitude is defined by
\be
  Im F_{N}(s,t) = \sigma_{tot}/(0.389 \cdot 4 \pi) exp(B/2 t),
\ee
 it is evident from  (\ref{rsq}) that the determination of
 the real part depends on
 $n, \sigma_{tot}, B$.
     The magnitude of $\sigma_{tot}$ determined from experimental
   data depends on the normalization parameter $n$ which reflects
   the experimental error in determining $d\sigma/dt$ from $dN/dt$.

   Let us examined this expression for the $pp$-scattering.
 For this aim, let us make a gedanken experiment and calculate
  $d\sigma/dt$
 with definite parameters taking them as the experimental points.
 In this case, we know exactly what we obtain at the end of our
 calculation. In this report we drop  the full analysis of this model
 experiment and dwell only on the special point.

  For the $pp$-scattering at high energies,
  the equation (\ref{rsq}) has a remarkable property.
 If we expand the expression under the radical sign, we obtain
\be
 (n-1)(Im F_{C} +Im F_{N})^2 + n (Re F_{C} +Re F_{N})^2 .
\ee


 As the real part of the Coulomb scattering amplitude is negative and
 the real part of the nuclon scattering amplitude is positive, it is clear
 that this expression  - $Del$ has a minimum situated on the scale
 of $t$  independent of $n$ and $\sigma_{tot}$ shown
 in Fig. 1 ( a and b).

 So, the position of the minimum  gives us
  $t_{er}$ where $Re F_{N} = -Re F_{C}$. As we know the
 Coulomb amplitude, we estimate the real part of the
 $pp$-scattering amplitude at this point. Note that all other methods give us
 the real part only in a sufficiently wide interval of the transfer
 momenta.
  This method works only in the case of the positive real part
 of the nucleon amplitude, and it is especially
  good in the case of large $\rho$. So, it is
 interesting for the future experiment on RHIC.

    Though in the range of ISR we have small $\rho(s, t \approx 0)$ and
 few experimental points, let us try to examine one experiment,
 for example, at $\sqrt{s}=52.8\ GeV$. This analysis is shown in Fig. 2.
 One can see that in this case the minimum is sufficiently large, and
 $-t_{min}= (3.3 \pm 0.1)10^{-2} \ GeV^2$.
 The corresponding real part
 equals $0.442 \pm 0.014 \ GeV$. Note that this magnitude is absolute
 and independent of the normalization coefficient and $\sigma_{tot}$
 If we take, as in the experiment, $\sigma_{tot}=42.38 \ mb$, we obtain
 $\rho=0.063 \pm 0.003$. Paper \ci{528} gives
 $\rho=0.077$.


    For the RHIC energies we can made a special random   procedure
 with the gaussian    distribution   of statistics
 and
  obtain, for example, the pictures for $\rho(t) = 0.135$ and
  for $\rho(t)= 0.175$  shown in Fig. 3 . The difference between these two
  representations is obvious. There is another interesting characteristic,
  the magnitude of second maximum.
   It is easy to  connect the size of the maximum with the magnitude of the
  real part of the scattering amplitude.
 After differentiated the representation of $Del$ we can determin the position
  of maximum of $Del = \Delta_{max} $
 and obtain:
\ba
  Re F_{N} (s,t) \approx \Delta_{max} ( 1 + \sqrt{\alpha B}
  \Delta_{max}^{-3/2}),
\ea
  where $B/2$ is the slope of $Re F_{N}(s,t)$.  It is to be noted that
  this representation
  slightly depends of our supposition on the form of $Re F_{N}(s,t)$.

  From Fig.3 we can look for the ratio of these two magnitudes of
  $Re F_{N}(s,t)$.
$ Re F_{N}(s,t)_1  =  1.35 $  and $Re F_{N}(s,t)_2 = 2.2$. These values give
  $\rho_1 = 0.13 $ (input $\rho_1 = 0.135$)  and
  $\rho_2 = 0.18 $ (input $\rho_2 = 0.175$).
$$ Re F_{N}(s,t)_1/Re F_{N}(s,t)_2 = 0.72 $$ and it can be compared with the
  input ratio of $\rho$:
 $$ \rho_1 / \rho_2 = .135 / .175 = 0.77 $$
   So, the magnitude of second maximum can give further information
   on the size of the real part of scattering amplitude.


    This point of $t_{er}$ is
 very important for
   the determination of the real part of spin-flip amplitude also.
 In the standard theory, the spin-flip hadronic amplitudes
 are expected to fall
 as $1/\sqrt{s}$ as $s\to\infty$, and they give negligible
  contributions at the
 present high energies. Only the spin-non-flip hadronic amplitudes
 $\Phi_1^{N}(s,t)\simeq \Phi_3^{N}(s,t)$ survive at these energies.
 So, the spin effects
 arise mostly from the interference between the
 non-flip hadron amplitudes with the spin-flip electromagnetic amplitudes.
  But there are some models which predict non-dying spin effects in the
  diffraction range at high energies \ci{gv-sf,ans}.
   The analyzing power is
\ba
   A_N = 4 \pi Im[ (\Phi_{1}+\Phi_{3})^{*} \Phi_{5}]/d\sigma/dt, \lab{an}
\ea
 or, separating the electromagnetic-hadron interference
  and pure hadron   parts, we obtain
\ba
- d\sigma/dt A_N /(4 \pi)  &=& Im F_{N} Re \Phi_{5}^{e}
   + Im F_{N} Re \Phi_{5}^{h} \\ \nonumber
 &-& ( Re F_{C} + Re F_{N}) Im \Phi_{5}^{h} .
\ea
  At the point  $t_{er}$, the real part of the hadron spin-flip
  amplitude is
\ba
 Re \Phi_{5}^{h} =   - A_N (d\sigma/dt)/(4 \pi Im F_{N})
         - \Phi_{5}^{e} .
\ea
  So, measuring the analyzing power at this point, we can obtain
  the magnitude of the real part of the hadron spin-flip amplitude
  without any theoretical assumption about its $t$-dependence.

  Note, that  $A_N d\sigma/dt = \sigma(++) - \sigma(+-)$. Hence
  the determination of the magnitude of the real part of hadron
  spin-non-flip amplitude contains only the difference of the cross sections
  with the parallel and antiparallel spins of nucleons.

 An the end, let us note  some additional method which can give
 the independent information on the real part of the hadron spin-non-flip
  amplitude using the measure of $A_N$ if the hadron spin-flip amplitude
  disapper at high energies \cite{gnsan}.
 Namely,
from eqs. (\ref{rsq}) and (\ref{an}), we get the real part of the hadron amplitude
\ba
  && Re F_{N}(s,t_i)= -Re F_{C}(s,t_i)   \nonumber    \\
  & & + [n / \pi d\sigma/ dt
   - (Im F_{C}(t_i)+ \widetilde A_N \ n d\sigma/dt)^2]^{1/2},
\ea
where
\ba
\widetilde A_N\equiv -A_N^{exp}(s,t_i)/(4\pi\cdot \Phi_{5}^{e}(t_i))\ .
\ea
 Again the real part of scattering amplitude is
  expressed in terms of the experimental quantities
$d\sigma^{exp}/dt$ and
$A_N^{exp}$, the normalisation factor $n$, and also
in terms of the theoretically
known electromagnetic amplitude $\Phi_{1}^{e}$.

The precise experimental measurements of $dN/dt$ and $A_N$ at
RHIC, as well as, if possible, at the Tevatron, will therefore give us
unavailable information on the hadron elastic scattering at small t. New
phenomena at high energies could be therefore detected without going through the
usual arbitrary assumptions (such as the exponential form) concerning the hadron
elastic scattering amplitudes.

     {\it Acknowledgement.} {\hspace {0.5cm} The author expresses his deep
gratitude to W. Gurin,  B. Nicolescu and B.S. Nurushev
for fruitful discussions.

\newpage
     Captions

FIG.1.
 The model calculations of the $del$
 for the
 $pp$-scattering
 at RHIC energy $\sqrt{s}=540 GeV$ on different $n$ \\
 a) with $\sigma_{tot}=63 \ mb$; b) with $\sigma_{tot}=62 \ mb$.

FIG.2.
  The calculation of $Del$ for the
$pp$-scattering
 using the experimental data at $\sqrt{s}=52.8 \ GeV$. The lines
 are the polynomial fit of the points calculated with experimental data
and with different $n$

FIG.3.
  The  calculation of $del$ for the model $pp$-scattering
 with a) $\rho_1 =0.135$ and b) $\rho_2 =0.175$
 The solid, short-dashed, and dotted lines are the
 theoretical curves for $rho_2 = 0.175$, $rho_1 = 0.135$, $rho_0 = 0.$
  respectively.

\newpage

\begin{figure}[!ht]
\vskip -1.5cm
\epsfysize=60mm
\centerline{\epsfbox{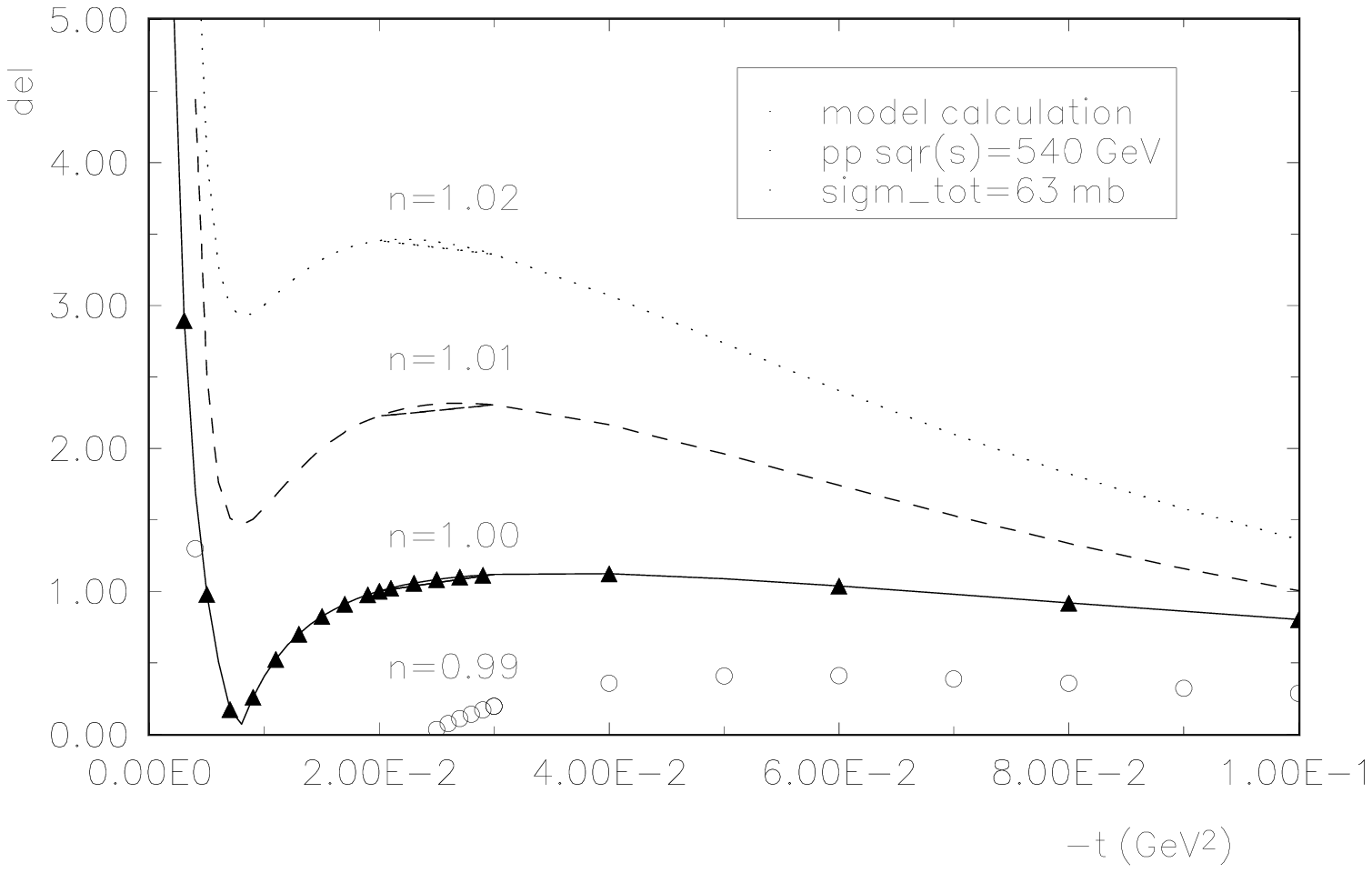}}
\vskip -1.cm
\epsfysize=60mm
\centerline{\epsfbox{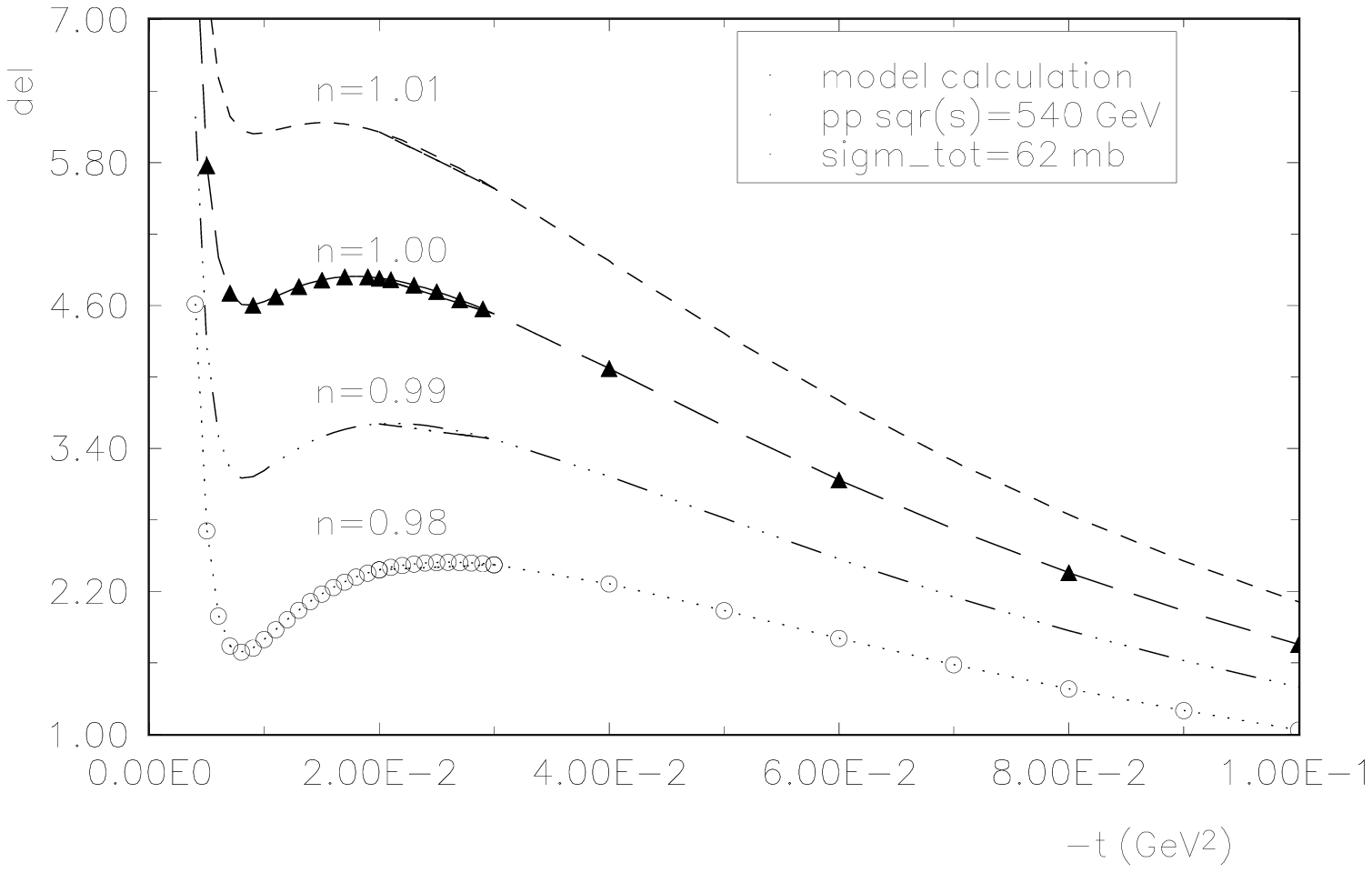}}
\vspace{-0.5cm}
\end{figure}

\newpage
\begin{figure}[!ht]
\epsfysize=60mm
\centerline{\epsfbox{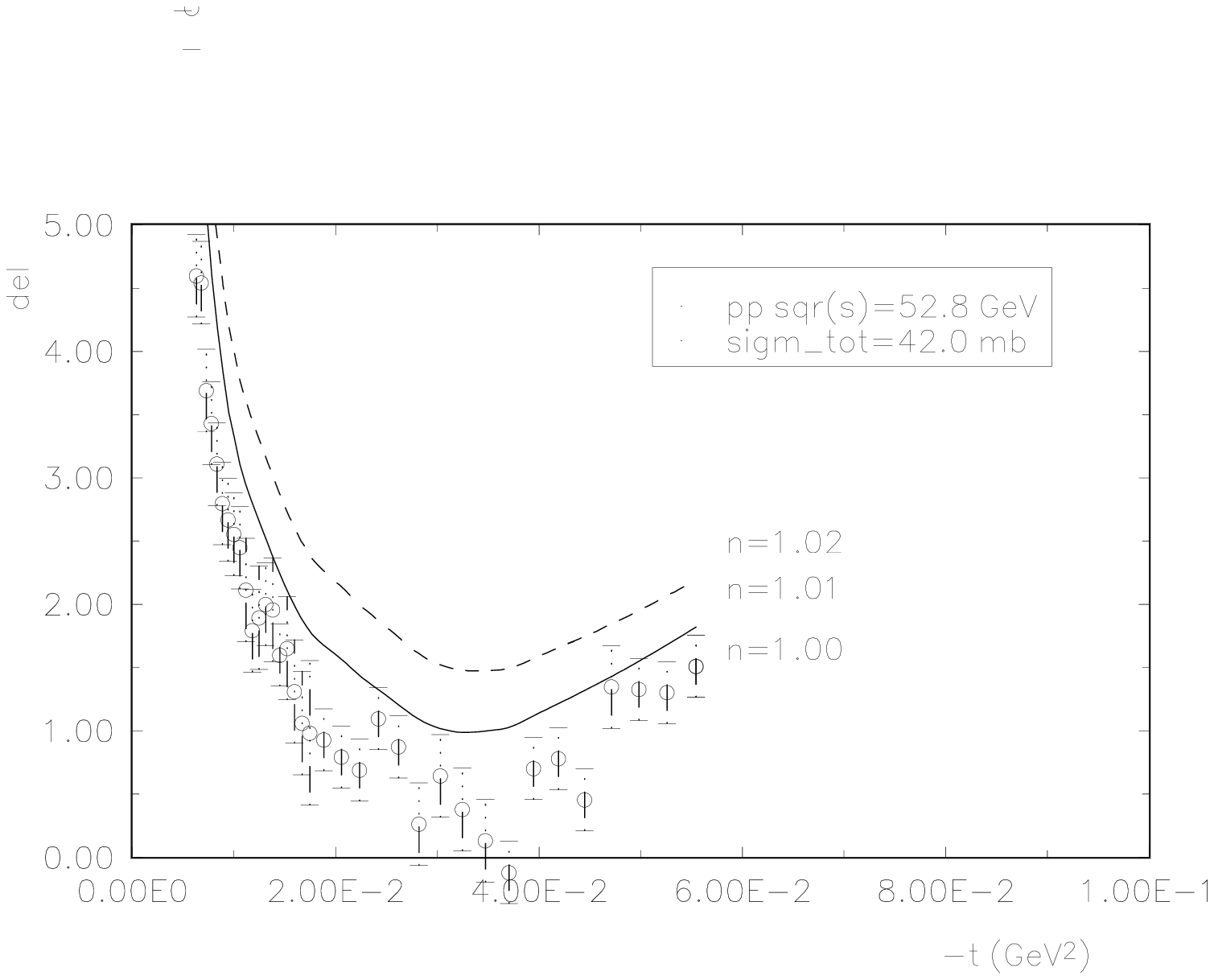}}
\vspace{0.5cm}
\end{figure}

\newpage

\begin{figure}[!ht]
\epsfysize=60mm
\centerline{\epsfbox{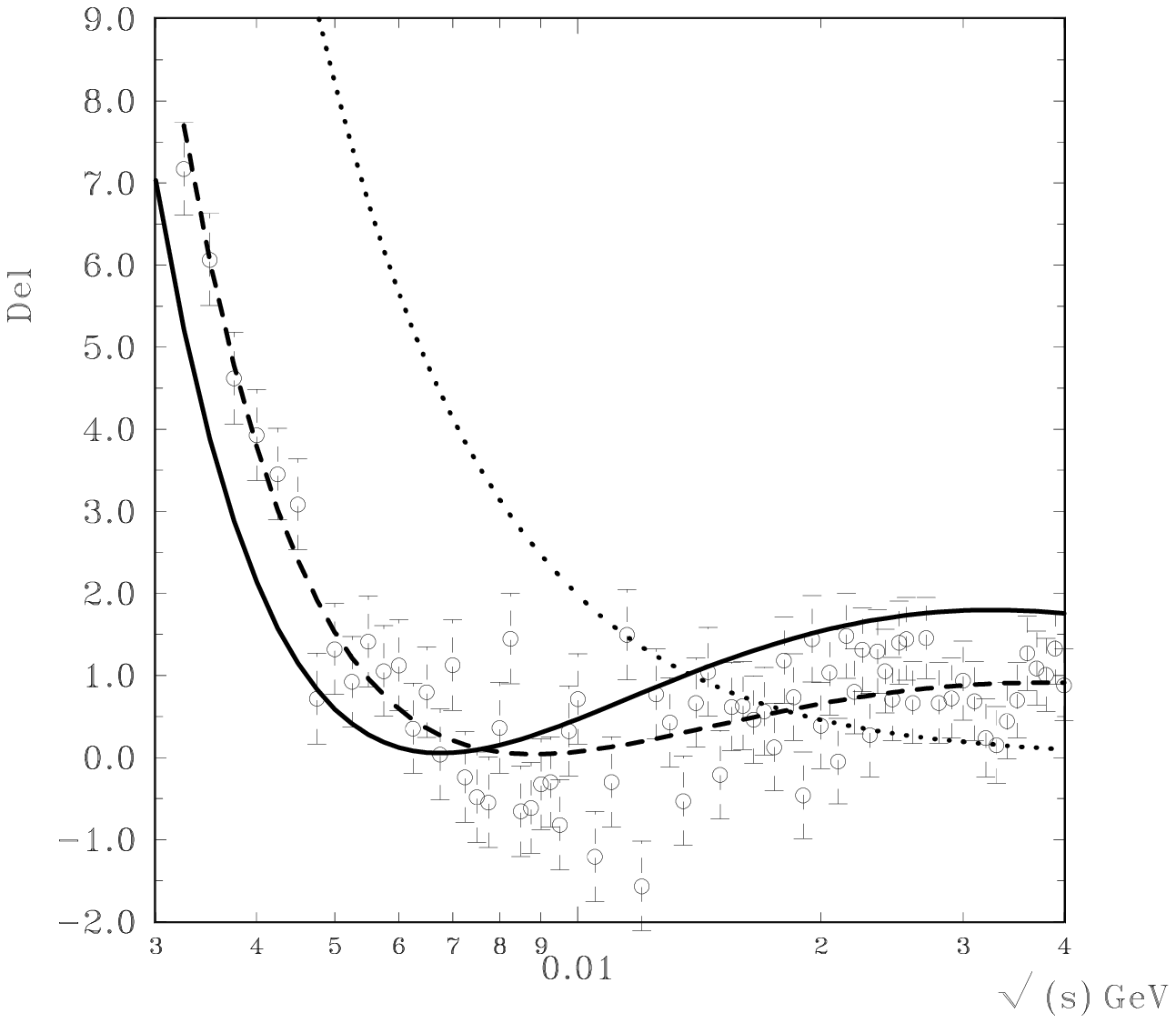}}
\epsfysize=60mm
\centerline{\epsfbox{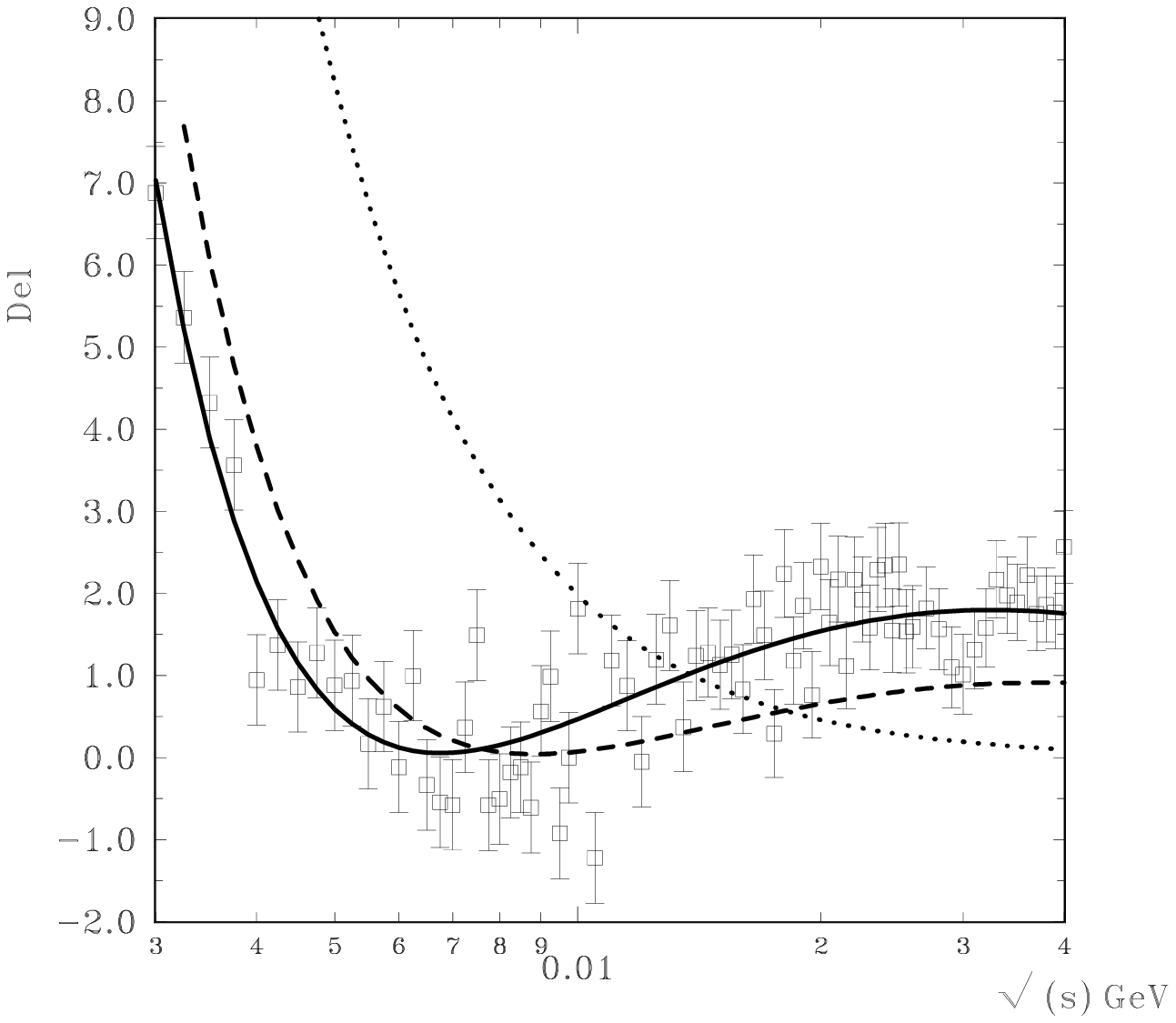}}
\vspace{-0.5cm}
\end{figure}

\end{document}